\documentclass[11pt]{article}
\newcommand{\Comment}[1]{{}}
\usepackage{amsfonts,amsthm,amsmath,amssymb,slashed}
\usepackage[textwidth = 430 pt, textheight = 630 pt]{geometry}
\usepackage{color} 
\usepackage{hyperref}
\usepackage[compat=1.1.0]{tikz-feynman}
\usepackage{comment}
\usepackage{subfigure}
\usepackage{placeins}

\Comment{\usepackage{color}
\definecolor{MyDarkBlue}{rgb}{0.15,0.15,0.45}
\usepackage[linktocpage=true]{hyperref}
\hypersetup{
colorlinks=true,
citecolor=MyDarkBlue,
linkcolor=MyDarkBlue,
urlcolor=MyDarkBlue,
pdfauthor={Matheus Cravo and Horatiu Nastase },
pdftitle={3-point function of $SO(3)$ global symmetry current in a three-dimensional toy model for holographic cosmology.},
pdfsubject={hep-th}
}

\usepackage[numbers,sort&compress]{natbib}
\usepackage{hypernat}}
\usepackage{graphicx}
\usepackage{cite}

\newcommand\ignore[1]{}
\def\one{{\,\hbox{1\kern-.8mm l}}}

\def\d{\partial}

\newcommand{\Cset}{{\,\,{{{^{_{\pmb{\mid}}}}\kern-.45em{\mathrm C}}}}}

\newcommand{\be}{\begin{equation}}
\newcommand{\bea}{\begin{eqnarray}}

\newcommand{\ee}{\end{equation}}
\newcommand{\eea}{\end{eqnarray}}

\begin{document}

\renewcommand{\thefootnote}{\fnsymbol{footnote}}

\makeatletter
\@addtoreset{equation}{section}
\makeatother
\renewcommand{\theequation}{\thesection.\arabic{equation}}

\rightline{}
\rightline{}




\begin{center}
{\LARGE \bf{\sc Holographic Cosmology $n$-point function map from the Wavefunction of the Universe}}
\end{center} 
 \vspace{1truecm}
\thispagestyle{empty} \centerline{
{\large \bf {\sc Matheus Cravo${}^{a}$}}\footnote{E-mail address: \Comment{\href{mailto:matheus.cravo@unesp.br}}
{\tt matheus.cravo@unesp.br}}
{\bf{\sc and}}
{\large \bf {\sc Horatiu Nastase${}^{a}$}}\footnote{E-mail address: \Comment{\href{mailto:horatiu.nastase@unesp.br}}
{\tt horatiu.nastase@unesp.br}}
                                                        }

\vspace{.5cm}


\centerline{{\it ${}^a$Instituto de F\'{i}sica Te\'{o}rica, UNESP-Universidade Estadual Paulista}} 
\centerline{{\it R. Dr. Bento T. Ferraz 271, Bl. II, Sao Paulo 01140-070, SP, Brazil}}
\vspace{.3cm}

\vspace{1truecm}

\thispagestyle{empty}

\centerline{\sc Abstract}

\vspace{.4truecm}

\begin{center}
\begin{minipage}[c]{380pt}
{\noindent In this note we show explicitly that, applying an extension of Maldacena's map for the wavefunction of the Universe
$Z[\phi]=\Psi[\phi]$ from de Sitter inflation to holographic cosmology, we find the relations previously derived for 
2-point functions, $\langle h_{ij}(p) h_{kl}(-p)\rangle \sim 1/[{\rm Im}\langle T_{ij}(-ip)T_{kl}(+ip)\rangle]$ and a similar one 
for currents, which were used to show that holographic cosmology matches CMBR data and solves Big Bang problems as 
well as inflation. Higher point functions are done similarly, and as an application, we check the result for the 
3-point functions of scalar and 
tensor fluctuations and find the result for 
the monopole non-Gaussianity arising from the 3-point functions of currents. The method is simple 
and potentially could be applied to calculate any observable in holographic cosmology.

}
\end{minipage}
\end{center}

\vspace{.5cm}

\setcounter{page}{0}
\setcounter{tocdepth}{2}

\newpage

\tableofcontents
\renewcommand{\thefootnote}{\arabic{footnote}}
\setcounter{footnote}{0}

\linespread{1.1}
\parskip 4pt



\section{Introduction}

The AdS/CFT correspondence \cite{Maldacena:1997re} was introduced as a holographic duality
relation between string theory in a gravitational 
background, at first of AdS type, and a conformal field theory, and it was soon extended to other gravitational backgrounds, 
as gauge/gravity duality (see the books  \cite{Nastase:2015wjb,Ammon:2015wua} for more information). But in 
most of these cases, the gravitational side was thought of as weakly coupled, and used to learn about a strongly coupled 
field theory side. But it was soon understood that (through many, many tests of the duality, some involving finite coupling), 
despite the duality being originally heuristically derived in the case of perturbative gravity, it is supposed to hold
generally, for any coupling. In particular, the relation between partition functions explicitly proposed by Witten \cite{Witten:1998qj}, 
and suggested also by GKP in \cite{Gubser:1998bc}, 
\be
Z_{\rm CFT}[\phi_0(\vec{x})]=Z_{\rm string}[\phi_0(\vec{x})]\;,\label{ZZAdSCFT}
\ee
where $\phi_0$ are viewed as sources for operators in the CFT and boundary conditions for fields in gravity, already 
suggested a more general application than its original one in \cite{Witten:1998qj} (to calculate strongly coupled 
operator $n$-point functions $\langle {\cal O}(\vec{x}_1)...{\cal O}(\vec{x}_n\rangle$ through differentiation with respect to 
$\phi_0(\vec{x})$, viewed as boundary value for the field $\phi$ in gravity, coupling to the operator ${\cal O}$ in the CFT). For 
instance, one simple example is the case of the Wilson loop, calculated in \cite{Maldacena:1998im,Rey:1998ik} through an 
extension of the meaning of the above partition function map. 

On the other hand, gravity in general was argued to be holographic, starting with the work of 't Hooft and Susskind
\cite{tHooft:1993dmi, Susskind:1994vu}, and was made concrete via the the AdS/CFT correspondence 
\cite{Maldacena:1997re}. One started to apply the holographic methods of AdS/CFT to regular, weakly 
coupled inflationary cosmology (which is a modification of de Sitter, sort of Wick rotated from Anti-de Sitter) in 
\cite{Witten:2001kn, Strominger:2001pn, Strominger:2001gp, Maldacena:2002vr} and the applications via strongly 
coupled QFTs were developed, for 
instance, in \cite{Maldacena:2011nz, Hartle:2012qb, Hartle:2012tv,Schalm:2012pi, Bzowski:2012ih, Mata:2012bx, Garriga:2013rpa, McFadden:2013ria, Ghosh:2014kba, Garriga:2014ema, Kundu:2014gxa, Garriga:2014fda, McFadden:2014nta, Arkani-Hamed:2015bza, Kundu:2015xta, Hertog:2015nia,Garriga:2015tea,  Garriga:2016poh,Hawking:2017wrd,Arkani-Hamed:2018kmz}.
In particular, in \cite{Maldacena:2002vr} it was proposed to use the formula, generalized from the basic AdS/CFT 
relation (\ref{ZZAdSCFT}),
\be
Z_{\rm CFT}[g^{(3)},\phi]=\Psi_{\rm gravity}[g^{(3)},\phi]\;,\label{ZPsi}
\ee
relating the CFT partition function $Z$ for a source $g^{(3)}$ for the Euclidean energy-momentum tensor and sources $\phi$ 
for other operators and the wave function of the Universe $\Psi$ in cosmology, 
for a 3-metric $g^{(3)}$ and other fields $\phi$ in gravity, in order to calculate inflationary $n$-point functions from CFT 
$n$-point functions. 

But the idea of holographic cosmology then emerged, where various authors understood various things by it, but
in particular McFadden and Skenderis 
\cite{McFadden:2009fg,McFadden:2010na} proposed that one should: 1) extend the previous ideas to include the more interesting 
case of {\em strongly coupled gravity} (a non-geometrical phase for cosmology) described through the use of 
{\em perturbative QFT}, thus finding the otherwise potentially difficult to calculate strongly coupled equivalent of inflation; and 2) 
since a top-down, string theoretic model is hard to obtain, consider a phenomenological approach for the QFT (similar 
to the phenomenological approach for condensed matter theory, or AdS/CMT, see the book \cite{Nastase:2017cxp} for more
details, just that now in the opposite regime for the coupling). Of course, that means that: 1) we cannot really talk about a 
"background gravitational solution" anymore; and 2) even if we did, it would generically not be of AdS or dS type. 
Fitting the free parameters to the CMBR data makes an equally good fit as the $\Lambda$CDM plus inflation one
\cite{Afshordi:2016dvb,Afshordi:2017ihr}, one solves the same Big Bang cosmology problems as well 
\cite{Nastase:2019rsn,Nastase:2020uon}, including (generically) the monopole problem \cite{Nastase:2020lvn} and
the cosmological constant problem becomes easier to understand \cite{Nastase:2018cbf}.
All of this suggests that indeed, the extension of the holography in cosmology ideas to strong coupling and phenomenological 
models is well deserved, and in particular we can use (\ref{ZPsi}) in these contexts as well. 
Indeed, in \cite{Bernardo:2018cow} it was used in an even more general context. The phenomenological holographic 
cosmology was further developed in
\cite{McFadden:2009fg, McFadden:2010na, McFadden:2010vh, McFadden:2011kk, Bzowski:2011ab, Coriano:2012hd, Kawai:2014vxa,McFadden:2010jw}, using methods from 
\cite{Skenderis:2002wp,Papadimitriou:2004ap, Papadimitriou:2004rz, Maldacena:2002vr}.

But, even though in \cite{McFadden:2009fg,McFadden:2010na}
one used more direct holographic methods in gravity to find the map from the $\langle T_{ij} T_{kl}\rangle$ 
correlators in the QFT to the cosmological $\langle h_{ij} h_{kl}\rangle$ correlators, it was claimed, for instance in 
\cite{McFadden:2010na,Afshordi:2016dvb,Afshordi:2017ihr} that a direct application of the Maldacena formula 
(\ref{ZPsi}), extended from de Sitter inflation to phenomenological holographic cosmology, leads to the same 
result. 

In this paper, we want to: 1) check this claim explicitly in all the relevant cases, since it is not immediately clear that the 
tensor structure and duality transformations don't introduce subtleties ; and 2) extend it to calculate, 
3-point functions of $h_{ij}$ and currents $j_i$, and apply to possible non-Gaussianities. 

The paper is organized as follows. In section 2 we derive the general formulas for correlators, and check that indeed, the 
2-point functions of scalar and tensor perturbations in cosmology are obtained from 2-point functions of 
energy-momentum tensors $T_{ij}$ in the quantum field theory by using the same formulas previously derived directly using
non-conformal holography. In section 3 we then apply the formulas to calculate the 2-point functions of gauge fields $A_i$ and 
relate it to the solution of the monopole problem in holographic cosmology, check that the 3-point functions of $h_{ij}$ 
are obtained from the 3-point functions of $T_{ij}$ and give non-Gaussianities, and calculate the 3-point function of $A_i$ 
and find the resulting monopole non-Gaussianities, and in section 4 we conclude.

\section{Holographic equations for cosmological $n$-point functions in the  wavefunction  approach.}
\label{wave_function_universe_approach}

\subsection{General formulas}

We start with the statement that the wavefunction of the Universe is equal to the partition function of a dual QFT,
\begin{equation}
    \Psi[h_{ij},\phi] = Z_{\text{QFT}}[h_{ij},\phi]\;,
\end{equation}
where we wrote $h_{ij}$ instead of $g^{(3)}_{ij}$ to emphasize that we want to calculate $n$-point 
functions of the spatial metric fluctuations, usually called $h_{ij}$.

We can reconstruct the wavefunction of the Universe, $\Psi$, which we will need to calculate the 
observable $n$-point functions, from correlators of the dual Euclidean 3-dimensional quantum field theory. 
Let $\Phi$ be a generic collection of bulk fields. Then we can write \cite{Maldacena:2002vr,McFadden:2010na}
\begin{equation}
    \Psi[\Phi]=\exp\left(\sum_{n=2}^{\infty}\frac{1}{n!}\int d^{3}x_{1}\ldots\int d^{3}x_{n}\langle\mathcal{O}^{\{i_{1}\}}(x_1) 
    \ldots \mathcal{O}^{\{i_{n}\}}(x_n)\rangle\Phi_{\{i_{1}\}}\ldots\Phi_{\{i_{n}\}}\right).\label{PsiPhi}
\end{equation}

Bulk correlators are obtained by integration of operators with the square of the wavefunction, as usual in quantum mechanics. 
For instance, the 2-point function of a bulk field $\Phi$ with a collection of indices $\{i\}$ is given by
\begin{equation}
    \langle\Phi_{\{i_{1}\}}(x)\Phi_{\{i_{2}\}}(y)\rangle=\int\mathcal{D}\Phi| \Psi|^{2}\Phi_{\{i_{1}\}}(x)\Phi_{\{i_{2}\}}(y).
\end{equation}

Taking the square of the wavefunction, we find 
\begin{equation}
    |\Psi|^2  = \exp\left\{ -\frac{1}{2}\int d^{3}x^{\prime}d^{3}y^{\prime} \Phi_{\{j_{1}\}}(x^{\prime}) [G_{\{j_1\}\{j_2\}}(x^{\prime},y^{\prime})]^{-1}\Phi_{\{j_{2}\}}(y^{\prime})+\ldots\right\} \;,
\end{equation}
where 
\begin{equation}
    [G_{\{j_1\}\{j_2\}}(x^{\prime},y^{\prime})]^{-1} \equiv  -2\text{Re} \langle\mathcal{O}^{\{j_{1}\}}(x^{\prime})\mathcal{O}^{\{j_{2}\}}\rangle .
\end{equation}

Therefore, the 2-point function of $\Phi$ is given by 
\begin{eqnarray}
    \langle\Phi_{\{i_{1}\}}(x)\Phi_{\{i_{2}\}}(y)\rangle &=& \frac{1}{\mathcal{N}} \int \mathcal{D}\Phi  \exp\left\{ -\frac{1}{2}
    \int d^{3}x^{\prime}d^{3}y^{\prime} \Phi_{\{j_{1}\}}(x^{\prime}) [G_{\{j_1\}\{j_2\}}(x^{\prime},y^{\prime})]^{-1}\times\right.\cr
    &&\left.\times \Phi_{\{j_{2}\}}(y^{\prime})+\ldots\right\}  \Phi_{\{i_{1}\}}(x)\Phi_{\{i_{2}\}}(y)\;, 
\end{eqnarray}
where $\mathcal{N}$ is the normalization of the wavefunction of the Universe. 
We can rewrite this expression in terms of functional derivatives of some auxiliar partition function $ \mathcal{Z}[\Phi,J]$,
\begin{equation}
    \langle\Phi_{\{i_{1}\}}(x)\Phi_{\{i_{2}\}}(y)\rangle = \frac{1}{\mathcal{N}} \left.\frac{\delta}{\delta J^{\{i_1\}}(x)}
    \frac{\delta}{\delta J^{\{ i_2\}}(y)}\mathcal{Z}[J]\right|_{J=0}, \label{2-point_function_from_functional_derivatives}
 \end{equation}
 where 
 \begin{eqnarray}
     \mathcal{Z}[J] &=& \frac{1}{\mathcal{N}} \int \mathcal{D}\Phi  \exp\left\{ -\frac{1}{2}\int d^{3}x^{\prime}
     d^{3}y^{\prime} \Phi_{\{j_{1}\}}(x^{\prime}) [G_{\{j_1\}\{j_2\}}(x^{\prime},y^{\prime})]^{-1}\Phi_{\{j_{2}\}}
     (y^{\prime})+\ldots \right. \nonumber \\
     && \hspace{6cm} \left. +\int d^{3}z^{\prime}J^{\{j_3\}}(z^{\prime})\Phi_{\{ j_3 \}}(z^{\prime})\right). \label{partition_function_0}
 \end{eqnarray}
 
If we consider {\em only the Gaussian part} of $\Psi$ above, 
we can define de analogous of a "generating functional for the free theory",
\begin{align}
    \mathcal{Z}_{0}[J] &= \frac{1}{\mathcal{N}} \int \mathcal{D}\Phi  \exp\left\{ -\frac{1}{2}\int d^{3}x^{\prime}d^{3}y^{\prime} 
    \Phi_{\{j_{1}\}}(x^{\prime}) [G_{\{j_1\}\{j_2\}}(x^{\prime},y^{\prime})]^{-1}\Phi_{\{j_{2}\}}(y^{\prime})\right. \nonumber \\
    & \hspace{6cm} \left. + \int d^3 z^\prime J^{\{ j_3\}}(z^\prime) \Phi_{\{ j_3 \}  } (z^\prime) \right\},
\end{align}
from which we can can write the Dyson's formula for the full "interacting theory", where by interactions we mean 
non-Gaussianities corresponding to the presence of $n$-point functions for $n\geq 3$,
\begin{eqnarray}
    \mathcal{Z}[J]
    &=&  \mathcal{Z}_0[J] +\frac{1}{3!}\int d^{3}x^{\prime}d^{3}y^{\prime}d^{3}z^{\prime}
    \lambda^{\{ j_1 \} \{ j_2 \} \{ j_3 \}}(x^{\prime},y^{\prime},z^{\prime})\nonumber \\
    &&  \hspace{3cm}\times  \frac{\delta}{\delta J^{\{ j_1 \}}(x^{\prime})}\frac{\delta}{\delta J^{\{ j_2 \}}
    (y^{\prime})}\frac{\delta}{\delta J^{\{ j_3 \}}(z^{\prime})} \mathcal{Z}_0[J] + ...\;, \label{generating_functional_dysons_formula}
\end{eqnarray}
where 
\begin{equation}
    \lambda^{\{ j_1 \} \{ j_2 \} \{ j_3 \}}(x^{\prime},y^{\prime},z^{\prime}) \equiv 2\text{Re}\langle\mathcal{O}^{\{j_{1}\}}
    (x^{\prime})\mathcal{O}^{\{j_{2}\}}(y^{\prime})\mathcal{O}^{\{j_{3}\}}(z^{\prime})\rangle. \label{3_vertex}
\end{equation}

The $\mathcal{Z}_0[J]$ generating functional can be exactly solved, since it is just a Gaussian integration,
\begin{equation}
    \mathcal{Z}_0[J] = \mathcal{N}\exp \frac{1}{2}\int d^{3}x^{\prime}d^{3}y^{\prime}J^{\{ j_1 \}}(x^{\prime}) 
    G_{\{ j_1\} \{ j_2\}}(x^\prime,y^\prime) J^{\{ j_2 \}}(y^{\prime}), \label{partition_function_1}
\end{equation}
where 
\begin{equation}
    G_{\{ j_1\} \{ j_2\}}(x^\prime,y^\prime) = -\frac{1}{2\text{Re} \langle\mathcal{O}^{\{j_{1}\}}(x^{\prime})
    \mathcal{O}^{\{j_{2}\}}(y^\prime)\rangle}. \label{definition_inverse_propagator}
\end{equation}

We can use \eqref{generating_functional_dysons_formula} to calculate the 3-point function of a bulk field $\Psi$ in terms 
of the 3-point function of a boundary operator $\mathcal{O}$. In this case, only the term with six functional derivatives 
on the auxiliary sources will contribute,
\begin{align}
    & \langle\Phi_{\{i_{1}\}}(x)\Phi_{\{i_{2}\}}(y)\Phi_{\{i_{3}\}}(z)\rangle 
     = \frac{1}{3!\mathcal{N} }\int d^{3}x^{\prime}d^{3}y^{\prime}d^{3}z^{\prime}\lambda^{\{j_{1}\}
     \{j_{2}\}\{j_{3}\}}(x^{\prime},y^{\prime},z^{\prime}) \nonumber \\ 
    &  \times\left\{ \frac{}{} G_{\{i_{3}\}\{j_{3}\}}(z,z^{\prime})G_{\{i_{2}\}\{j_{2}\}}(y,y^{\prime})G_{\{i_{1}\}\{j_{1}\}}
    (x,x^{\prime})+G_{\{i_{3}\}\{j_{3}\}}(z,z^{\prime})G_{\{i_{1}\}\{j_{2}\}}(x,y^{\prime})G_{\{i_{2}\}\{j_{1}\}}(y,x^{\prime}) 
    \right. \nonumber \\ 
    & +G_{\{i_{2}\}\{j_{3}\}}(y,z^{\prime})G_{\{i_{3}\}\{j_{2}\}}(z,y^{\prime})G_{\{i_{1}\}\{j_{1}\}}(x,x^{\prime})
    +G_{\{i_{1}\}\{j_{3}\}}(x,z^{\prime})G_{\{i_{3}\}\{j_{2}\}}(z,y^{\prime})G_{\{i_{2}\}\{j_{1}\}}(y,x^{\prime}) \nonumber \\ 
    & \left.+ G_{\{i_{2}\}\{j_{3}\}}(y,z^{\prime})G_{\{i_{1}\}\{j_{2}\}}(x,y^{\prime})G_{\{i_{3}\}\{j_{1}\}}(z,x^{\prime})
    +G_{\{i_{1}\}\{j_{3}\}}(x,z^{\prime})G_{\{i_{2}\}\{j_{2}\}}(y,y^{\prime})G_{\{i_{3}\}\{j_{1}\}}(z,x^{\prime}) \frac{}{} \right. \nonumber \\
    & \left.   +  \text{(disconnected diagrams) } \frac{}{}\right\} \mathcal{Z}_0[ J=0] . \label{3-point_function_x_space}
\end{align}

Diagramatically, these terms are just all possible ways to connect three external points $(x,y,z)$  to three internal points 
$(x^\prime, y^\prime, z^\prime)$ which are integrated over all space. The role of "vertex" is played by the 3-point 
function of the boundary operator $\langle \mathcal{O} \mathcal{O}\mathcal{O} \rangle$ as defined in 
\eqref{3_vertex}, with the "propagators" being $\langle \mathcal{O} \mathcal{O} \rangle$.

The disconnected diagrams are analogous to “bubble” diagrams that appears in the expansion of the generating function 
of the interacting theory. These terms will not contribute to the 3-point functions, since they can be eliminated by 
computing the normalization of the wavefunction, or equivalenty, by setting $\mathcal{Z}[J=0] = 1$. The only terms we 
need to consider are the connected ones, and each of these terms will contribute to 
$\langle \Phi_{\{i_{1}\}}(x)\Phi_{\{i_{2}\}}(y)\Phi_{\{i_{3}\}}(z)\rangle $, with
\begin{equation}
    \frac{2\text{Re}\langle\mathcal{O}^{\{j_{1}\}}(p_{1})\mathcal{O}^{\{j_{2}\}}(p_{2})\mathcal{O}^{\{j_{3}\}}(p_{3})\rangle}
    {\left[-2\text{Re}\langle \mathcal{O}^{\{i_{1}\}}(p_{1})\mathcal{O}^{\{j_{1}\}}(-p_{1})\rangle\right]\left[-2\text{Re}\langle 
    \mathcal{O}^{\{i_{2}\}}(p_{2})\mathcal{O}^{\{j_{2}\}}(-p_{2})\rangle\right]\left[-2\text{Re}\langle \mathcal{O}^{\{i_{3}\}}
    (p_{3})\mathcal{O}^{\{j_{3}\}}(-p_{3})\rangle\right]},
\end{equation}
after renaming dummy indices and taking the Fourier transform. Hence, we have a factor of 6 since we have 6 diagrams giving 
the same contribution, which cancels the factor $1/3!$ in \eqref{3-point_function_x_space}. In momentum space, the final answer is
\begin{align}
    &\langle\Phi_{\{i_{1}\}}(p_1)\Phi_{\{i_{2}\}}(p_2)\Phi_{\{i_{3}\}}(p_3)\rangle 
    = 2\text{Re}\langle\mathcal{O}^{\{j_{1}\}}(p_{1})\mathcal{O}^{\{j_{2}\}}(p_{2})\mathcal{O}^{\{j_{3}\}}(p_{3})\rangle \nonumber \\
     &  \times \frac{1}{\left[-2\text{Re}\langle \mathcal{O}^{\{i_{1}\}}(p_{1})\mathcal{O}^{\{j_{1}\}}(-p_{1})\rangle\right]
     \left[-2\text{Re}\langle \mathcal{O}^{\{i_{2}\}}(p_{2})\mathcal{O}^{\{j_{2}\}}(-p_{2})\rangle\right]\left[-2\text{Re}
     \langle \mathcal{O}^{\{i_{3}\}}(p_{3})\mathcal{O}^{\{j_{3}\}}(-p_{3})\rangle\right]}.\label{3pointgeneral}
\end{align}

This matches the result in the case of perturbative, de Sitter inflation, obtained by Maldacena in \cite{Maldacena:2002vr}.

In a completely similar manner, for the 4-point function we obtain 
\bea
&&\langle \Phi_{\{i_1\}}(p_1)\Phi_{\{i_2\}}(p_2)\Phi_{\{i_3\}}(p_3)\Phi_{\{i_4\}}(p_4)\rangle=\cr
&&\frac{1}{\left[-2{\rm Re}\langle {\cal O}^{\{i_1\}}
(p_1){\cal O}^{\{j_1\}}(-p_1)\right]\left[-2\text{Re}\langle \mathcal{O}^{\{i_{2}\}}(p_{2})\mathcal{O}^{\{j_{2}\}}(-p_{2})\rangle\right]}
\times\cr
&&\times\frac{1}{\left[-2\text{Re}\langle \mathcal{O}^{\{i_{3}\}}(p_{3})\mathcal{O}^{\{j_{3}\}}(-p_{3})\rangle\right]\left[-2\text{Re}
     \langle \mathcal{O}^{\{i_{4}\}}(p_4)\mathcal{O}^{\{j_{4}\}}(-p_{4})\rangle\right]}\times\cr
&&\times\left\{\left[2{\rm Re}\langle {\cal O}^{\{j_1\}}(p_1){\cal O}^{\{j_2\}}(p_2){\cal O}^{\{j_3\}}(p_3){\cal O}^{\{j_4\}}(p_4)\rangle
\right]+\right.\cr
&&\left.+\frac{1}{8}\left[2{\rm Re}\langle{\cal O}^{\{j_1\}}(p_1){\cal O}^{\{j_2\}}(p_2){\cal O}^{\{j\}}(p_1+p_2)\rangle\right]
\left[2{\rm Re}\langle{\cal O}^{\{j\}}(p_1+p_2){\cal O}^{\{j_3\}}(p_3){\cal O}^{\{j_4\}}(p_4)\rangle\right]\right.\cr
&&\left.+\frac{1}{8}\left[2{\rm Re}\langle{\cal O}^{\{j_1\}}(p_1){\cal O}^{\{j_3\}}(p_3){\cal O}^{\{j\}}(p_1+p_3)\rangle\right]
\left[2{\rm Re}\langle{\cal O}^{\{j\}}(p_1+p_3){\cal O}^{\{j_2\}}(p_2){\cal O}^{\{j_4\}}(p_4)\rangle\right]\right.\cr
&&\left.+\frac{1}{8}\left[2{\rm Re}\langle{\cal O}^{\{j_1\}}(p_1){\cal O}^{\{j_4\}}(p_4){\cal O}^{\{j\}}(p_1+p_4)\rangle\right]
\left[2{\rm Re}\langle{\cal O}^{\{j\}}(p_1+p_4){\cal O}^{\{j_2\}}(p_2){\cal O}^{\{j_3\}}(p_3)\rangle\right]\right\}\;,\cr
&&
\eea
where we see that again
\be
2{\rm Re}\langle {\cal O}^{\{ j_1\}}(p_1){\cal O}^{\{j_2\}}(p_2){\cal O}^{\{j_3\}}(p_3){\cal O}^{\{j_4\}} (p_4)\rangle
\equiv \lambda ^{\{j_1\}\{j_2\}\{j_3\}\{j_4\}}(p_1,p_2,p_3,p_4)
\ee
acts as a 4-point vertex, 1/8 is a symmetry factor for the Feynman diagram, and the sum is in the $\{$ $\}$ brackets is
over the 4 Feynman diagrams. 

The formula can be then easily generalized to find any $n$-point function, by summing over the tree  Feynman diagrams 
made using the "propagators" and "vertices". Then any observable, obtained from fluctuations of the fields, can be 
found from these $n$-point functions, and thus obtained from the set of $n$-point functions of the quantum field theory 
(where the real parts of correlators of ${\cal O}^{\{j\}}(p)$, $2{\rm Re}\langle {\cal O}^{\{j_1\}}(p_1)...{\cal O}^{\{j_n\}}(p_n)\rangle$, 
act as vertices).

\subsection{Matching holographic cosmology to the CMBR using the domain-wall/cosmology correspondence}

Consider as the field the spatial 3-metric, $\Psi_{\{i_1\}} = h_{ij}$. 
The dual operator is the energy-momentum tensor $\mathcal{O}^{\{ i_1 \}} = T^{ij}$, 
such that the 2-point function of the spatial metric is given by 
\begin{equation}
     \langle h_{ij}(x) h_{kl}(y) \rangle = \frac{1}{\mathcal{N}} \left.\frac{\delta}{\delta J^{ij}(x)}\frac{\delta}
     {\delta J^{kl}(y)}\mathcal{Z}[J_{mn}]\right|_{J_{mn}=0}.
\end{equation}

Note that the result of any odd number of $\delta/\delta J$ derivatives acting on $\mathcal{Z}_0$ is zero 
after taking $J=0$. Hence, for the 2-point function, the first contribution (analogous to a tree-level result) comes only 
from the Gaussian part of the wavefunction, which gives in momentum space
\begin{equation}
    \langle h_{ij}(p) h_{kl}(-p)\rangle=-\frac{1}{2\text{Re}\langle T^{ij}(p)T^{kl}(-p)\rangle}\;,
     \label{2-point_function_graviton_momentum_space}
\end{equation}
where the delta function for conservation of the momenta was omitted. 

The scalar and tensor power spectra are defined in cosmology respectively as 
\begin{equation}
    \Delta_S^2 (q) =  \frac{q^3}{2\pi^2} \langle \xi(q) \xi(-q) \rangle, \quad \quad \Delta^2_T (q) 
    = \frac{q^3}{2\pi^2} \langle \gamma_{ij}(q) \gamma_{ij}(-q) \rangle,
\end{equation}
where $\xi$ is the comoving curvature and $\gamma_{ij}$ is the transverse traceless metric perturbation. 
Following the prescription in \cite{Maldacena:2002vr}, $\xi$ couples to the trace 
of the energy-momentum tensor ${T^i}_i$ while $\gamma_{ij}$ couples to the transverse traceless part of the 
energy-momentum tensor $T_{ij}^\perp$. For the scalar power spectrum, this means that
\begin{equation}
    \Delta_S^2(q) = - \frac{q^2}{4\pi^2} \frac{1}{\text{Re} \langle  T(q) T(-q) \rangle}.
\end{equation}

For the tensor power-spectrum, it is usefull to work with the helicity basis, where $\gamma_{ij} 
= \gamma(q)^s \epsilon_{ij}^s(q)$ and $s=+,-$. The polarization vector satisfies
\begin{equation}
  \epsilon_{ij}(q)^s \epsilon_{ij}(-q)^{s^\prime}= 2\delta^{ss^\prime}\;,
\end{equation}
and the energy-momentum tensor in the helicity basis is given by 
\begin{equation}
  T^s(q) = \frac{1}{2}\epsilon_{ij}^s(-q) T^{ij} (q). \label{T_helicity_basis}
\end{equation}

Therefore, the tensor power-spectrum in helicity basis is given by
\begin{align}
  \Delta_T^2 &= \frac{q^3}{\pi^2} \sum_{s} \langle \gamma(q)^s \gamma(-q)^{s} \rangle, \nonumber \\
  &= \frac{q^2}{\pi^2} \sum_s \left[ -\frac{1}{2\text{Re}  \langle T^s(q) T^s(-q) \rangle} 
   \right]. \label{tensor_power_spectrum_from_wave_functional_helicity_basis}
\end{align}

The 2-point function of the energy-momentum tensor in the theory dual to cosmology can 
be decomposed into the two types of transverse tensor structures (the $T_{ij}$ correlators are transverse because of
gauge invariance)
\begin{equation}
     \langle T^{ij}(q)T^{kl}(- q) \rangle = \Pi^{ijkl} A(q) + \pi^{ij}\pi^{kl}  B(q)\;,
\end{equation}
where the two types of transverse projectors are
\begin{equation}
     \Pi^{ijkl} =\frac{1}{2}\left(\pi^{ik}\pi^{lj}+\pi^{il}\pi^{kj}-\pi^{ij}\pi^{kl}\right), \hspace{2cm} \pi^{ij}=\delta^{ij}-\frac{q^{i}q^{j}}{q^{2}}.
\end{equation}

Then, for the scalar power spectrum, we are interested in the quantity
\begin{equation}
  \langle T(q)T(- q) \rangle = \Pi^{iikk} A(q) + \pi^{ii}\pi^{kk}  B(q).
\end{equation}

But since $\pi^{ii} =2$ and $\pi^{ij}\pi^{ji} = 2$, we have
\begin{equation}
    \langle T(q)T(- q) \rangle = 4 B(p) ,
\end{equation}
such that 
\begin{equation}
    \Delta_S^2(q) = -\frac{q^3}{16\pi^2}\frac{1}{\text{Re} B(q)}. \label{scalar_power_spectrum_wavefunction_approach}
\end{equation}

For the tensor power spectrum, $\langle T^s(q) T^s(-q) \rangle$ contains only the transverse traceless part of the 2-point function:
\begin{equation}
  2 \langle T^s(q) T^s(-q) \rangle = A(q), \label{2-point_function_TT_helicity_basis}
\end{equation}
where there is no summation over $s$, meaning that this expression is valid both for $s=+$ and $s=-$. Substituting \eqref{2-point_function_TT_helicity_basis} in \eqref{tensor_power_spectrum_from_wave_functional_helicity_basis}, we  find
\begin{equation}
  \Delta_T^2 (q) = -\frac{2q^3}{\pi^2} \frac{1}{\text{Re} A(q)},
\end{equation}
where the factor of $2$ comes from the summation over $s$. All in all, the 2-point functions from the wavefunction approach,
to be compared with the CMBR data, are 
\begin{equation}
  \Delta_S^2(q) = -\frac{q^3}{16\pi^2}\frac{1}{\text{Re} B(q)}, \hspace{2cm} \Delta_T^2 (q) 
  = -\frac{2q^3}{\pi^2} \frac{1}{\text{Re} A(q)}. \label{scalar_power_spectrum_wavefunction_approach_final}
\end{equation}

These are the expressions relating the scalar and tensor power spectrum in cosmology to the coefficients 
$A(q)$ and $B(q)$ of the decomposition of the energy-momentum tensor dual to cosmology. This result is 
equivalent to holographic phenomenological approach to cosmology proposed by McFadden and Skenderis, 
where the dual field theory to cosmology is a \emph{pseudo-QFT}, obtained by analytical continuation of results 
calculated using the domain wall/cosmology correspondence \cite{McFadden:2009fg,McFadden:2010na}. Indeed, in this latter
case, the scalar and tensor power spectrum are given by
\begin{equation}
    \Delta_S^2(q) = -\frac{q^3}{16\pi^2}\frac{1}{\text{Im} B(-iq)}, \hspace*{1cm} \Delta_T^2 (q) 
    = -\frac{2q^3}{\pi^2} \frac{1}{\text{Im} A(-iq)}.
\end{equation}

By direct comparison of these equations, we conclude that both results are the same if  
we use the domain wall/cosmology correspondence map,
\begin{equation}
  \text{Re} A(q) = \text{Im} A_D(-iq),\hspace{1cm} \text{Re} B(q) = \text{Im} B_D(-iq),
\end{equation}
or, in terms of the two-point function of the energy-momentum tensor,
\begin{equation}
  \text{Re} \langle T^{ij}(q) T^{kl}(-q) \rangle = \text{Im} \langle T_D^{ij}(\bar q) T_D^{kl}(-\bar q) \rangle \label{final_result},
\end{equation}
where the subscript $D$ was used to indicate quantities obtained in the domain wall.

\section{Applications}

\subsection{The 2-point function of vectors and the solution to the monopole problem}

For the solution of the monopole problem of Big Bang cosmology, it is necessary that the monopole distribution 
is diluted during the non-geometric phase of holographic cosmology, or that the monopole field 2-point function 
decays in cosmological time, dual to inverse RG flow. In quantum field theory, that corresponds to 
having the monopole current be a relevant operator, increasing in the IR, so decreasing in the UV, or 
that the global symmetry current be an irrelevant operator, increasing in the UV
\cite{Nastase:2019rsn,Nastase:2020uon}.

We are then interested in the 2-point functions of (transverse) gauge fields, so 
for a bulk gauge field $A_\mu$ dual to some boundary current $j^\mu$, in which case we obtain
\begin{equation}
    \langle A^i(p) A^j(-p) \rangle = - \frac{1}{2} \frac{1}{\text{Re}\langle j_i(p) j_j(-p) \rangle}.\label{AAjj}
\end{equation}

As discussed in \cite{Nastase:2019rsn,Nastase:2020uon}, the generalized conformal structure fixes the form of the 
correlator $\langle j_i(p) j_j(-p) \rangle$ like in the pure conformal case \cite{Witten:2003ya,Herzog:2007ij}, to be 
\begin{equation}
    \langle j_i(p) j_j(-p) \rangle = \left(p^2 \delta_{i j}-p_i p_j\right) \frac{t}{2 \pi \sqrt{p^2}}+\epsilon_{i j k} p_k \frac{w}{2 \pi}\;,
\end{equation}
where $t$ and $w$ are coefficient functions of the coupling fully determined by the theory. In particular, for the 
parity invariant class of models we are interested in holographic cosmology, the term proportional to $w$ is zero, 
such that we have just 
\begin{equation}
    \langle j_i(p) j_j(-p) \rangle = \left(p^2 \delta_{i j}-p_i p_j\right) \frac{t}{2 \pi \sqrt{p^2}}.
\end{equation}

But it was found \cite{Nastase:2019rsn,Nastase:2020uon,Nastase:2020lvn}
that generically the current $j_i$ is marginally irrelevant operator (with $\delta(j)>0$), 
\be
t\propto p^{2\delta(j)},\;\;\; \delta(j)>0.
\ee

Finally, since under the electric-magnetic duality the action of S-duality takes $t\rightarrow t^{-1}$, the 
{\em monopole} current field 2-point function is 
\begin{equation}
    \langle \tilde j_i(p) \tilde j_j(-p) \rangle = \left(p^2 \delta_{i j}-p_i p_j\right) \frac{1}{2 \pi t \sqrt{p^2}}\propto p^{1-2\delta}\;,
\end{equation}
so the monopole current $\tilde j_i$ is marginally relevant (with $\delta(\tilde j)=-\delta(j)<0$), so the 
monopole current 2-point function decreases in the UV. 

We note that the current 2-point function is real if $t$ is real, so we can just put the real part on $t$, and drop it elsewhere. 
In order to go to cosmology, we also need to do the Wick rotation $p\rightarrow \bar p=-ip$.

Then the corresponding correlation function for the bulk gauge field is 
\be
    \langle A^{i}(p)A^{j}(-p)\rangle =  -\pi\frac{p}{\left(p^{2}\delta_{ij}-p_{i}p_{j}\right)}\frac{1}{{\rm Re}\; t(p)}
    =+\pi\frac{\bar p}{\bar p^2\delta_{ij}
    -p\bar p_i\bar p_j}\frac{1}{{\rm Im} \; t(\bar p)}\;,
\ee
or, by contraction with $\delta_{ij}$, we can write 
\bea
\langle A^i (p)A_i (-p)\rangle&=&-\frac{\pi}{(d-1)p}\frac{1}{{\rm Re}\; t}=-\frac{1}{2{\rm Re} 
\langle j_i(p)j^i(-p)\rangle}=-\frac{2\pi^2}{(d-1)^2p^2}{\rm Re}\langle \tilde j_i(p)\tilde j^i(-p)\rangle\cr
&=&\frac{2\pi^2}{(d-1)^2\bar p^2}{\rm Im}\langle \tilde j_i(\bar p)\tilde j^i(-\bar p)\rangle.
\eea

Thus the 2-point function of (classical) fluctuations of the gauge field $A^i$, which means monopoles, 
decreases in the UV (mapped to the future in cosmology), like $t^{2\delta(\tilde j)}=t^{-2\delta(j)}$. 

%
%

\subsection{Non-Gaussianities in the CMBR tensor power spectrum}

The non-Gaussianities in the CMBR power spectrum are obtained from the $h_{ij}$ 3-point function. According to the general 
formula (\ref{3pointgeneral}), we have
\begin{align}
    \langle h_{kl}(p_{1})h_{ij}(p_{2})h_{gh}(p_{3}) \rangle &= 2\text{Re}\langle T^{ef}(p_{1})
    T^{cd}(p_{2})T^{ab}(p_{3})\rangle \nonumber \\
     & \hspace{-3cm} \times \frac{1}{\left[-2\text{Re}\langle T^{gh}(p_{3})T^{ab}(-p_{3})\rangle\right]
     \left[-2\text{Re}\langle T^{ij}(p_{2})T^{cd}(-p_{2})\rangle\right]\left[-2\text{Re}\langle T^{kl}(p_{1})T^{ef}(-p_{1})\rangle\right]}.
\end{align}

Then one must do the Wick rotation to cosmology as in (\ref{final_result}). 

This is indeed what was used in \cite{McFadden:2010vh,McFadden:2011kk,Bzowski:2011ab} to calculate the 
non-Gaussianities in holographic cosmology, formula derived also in the alternative way, directly in non-conformal 
holography.


\subsection{Monopole non-Gaussianity from the 3-point function of bulk gauge field.}

In \cite{Cravo:2023fqf}, 
we have calculated the 3-point function of global symmetry currents $j_\mu^a$, and found that, at one-loop,
\begin{align}
    \langle j_{\mu}^a(p_{1})j_{\nu}^b(-p_{2})j_{\rho}^c(-p_{3})\rangle &= N^2\epsilon^{abc} c_{0} \left[p_{1\mu}\left(c_{1}
    p_{2\nu}p_{2\rho}+c_{2}p_{3\nu}p_{3\rho}+c_{3}p_{3\nu}p_{2\rho}+c_{4}p_{3\rho}p_{2\nu}\right)\right. \nonumber \\
    &\quad\quad\quad  +p_{2\mu}\left(c_{5}p_{2\nu}p_{2\rho}+c_{6}p_{3\nu}p_{3\rho}
    +c_{7}p_{3\nu}p_{2\rho}+c_{8}p_{3\rho}p_{2\nu}\right) \nonumber \\
    &\quad\quad\quad  +p_{3\mu}\left(c_{9}p_{2\nu}p_{2\rho}+c_{10}p_{3\nu}p_{3\rho}
    +c_{11}p_{3\nu}p_{2\rho}+c_{12}p_{3\rho}p_{2\nu}\right) \nonumber \\
    &\quad\quad\quad  +\delta_{\nu\rho}\left(c_{13}p_{1\mu}+c_{14}p_{2\mu}+c_{15}p_{3\mu}\right) \nonumber \\
    & \left. \quad\quad\quad  +\delta_{\mu\rho}\left(c_{16}p_{2\nu}+c_{17}p_{3\nu}\right) 
    +\delta_{\mu\nu}\left(c_{18}p_{2\rho}+c_{19}p_{3\rho}\right)\right], \label{three_point_function_coefficients}
\end{align}
where
\begin{align}
    c_{0} & = \frac{1}{4p_{1}p_{2}p_{3}\left(p_{1}+p_{2}+p_{3}\right)^{3}} \nonumber \\ 
    c_{1} & = -p_{3}\left[4p_{3}^{2}+\left(p_{1}+p_{2}\right){}^{2}+3p_{3}\left(p_{1}+p_{2}\right)\right], & 
    c_{2} & = p_{2}p_{3}\left(p_{1}-p_{2}+p_{3}\right), \nonumber \\
    c_{3} & = p_{2}p_{3}\left(p_{1}+p_{2}+3p_{3}\right), &
    c_{4} & = -p_{3}\left[2p_{3}^{2}+\left(2p_{1}+p_{2}\right)p_{3}+p_{2}\left(p_{1}+p_{2}\right)\right] \nonumber \\ 
    c_{5} & = 2p_{3}^{2}\left(p_{1}+p_{2}+p_{3}\right),& 
    c_{6} & = 2p_{2}^{2}\left(p_{1}+p_{2}+p_{3}\right) \nonumber \\ 
    c_{7} & = -2p_{2}p_{3}\left(p_{1}+p_{2}+p_{3}\right). & 
    c_{8} & = -\left(p_{1}+p_{2}+p_{3}\right)\left(p_{1}^{2}-p_{2}^{2}-p_{3}^{2}\right) \nonumber \\ 
    c_{9} & = p_{3}\left[p_{1}^{2}+\left(p_{2}+p_{3}\right)p_{1}+2p_{3}\left(p_{2}+p_{3}\right)\right], & 
    c_{10} & = p_{2}\left[p_{1}^{2}+\left(p_{2}+p_{3}\right)p_{1}+2p_{2}\left(p_{2}+p_{3}\right)\right] \nonumber \\ 
    c_{11} & = -2p_{2}p_{3}\left(p_{2}+p_{3}\right),& 
    c_{12} & = p_{1}^{3}+\left(p_{2}+p_{3}\right)\left[p_{1}^{2}+p_{2}^{2}+p_{3}^{2}+\left(p_{2}+p_{3}\right)p_{1}\right] \nonumber \\ 
    c_{13} & = -p_{1}p_{2}p_{3}\left(p_{1}+p_{2}+p_{3}\right)\left(p_{1}+p_{2}+2p_{3}\right),& 
    c_{14} & = 2p_{1}p_{2}p_{3}\left(p_{1}+p_{2}+p_{3}\right){}^{2} \nonumber \\ 
    c_{15} & = p_{1}p_{2}p_{3}\left(p_{2}+p_{3}\right)\left(p_{1}+p_{2}+p_{3}\right),& 
    c_{16} & = p_{1}p_{2}\left(p_{1}+p_{2}\right)p_{3}\left(p_{1}+p_{2}+p_{3}\right) \nonumber \\ 
    c_{17} & = p_{1}p_{2}p_{3}\left(p_{1}+p_{2}+p_{3}\right)\left(p_{1}+2p_{2}+p_{3}\right),& 
    c_{18} & = -p_{1}p_{2}p_{3}\left(p_{1}+p_{2}+p_{3}\right)\left(p_{1}+p_{2}+2p_{3}\right) \nonumber \\ 
    c_{19} & = -p_{1}p_{2}p_{3}\left(p_{1}+p_{3}\right)\left(p_{1}+p_{2}+p_{3}\right). \label{coefficients_final_case}
\end{align}

Moreover, the 2-point function at 2-loops was found to be \cite{Nastase:2019rsn,Nastase:2020uon}
\begin{equation}
\langle j^a_\mu(p)  j^b_\nu(-p) \rangle =  N^{2}\frac{p}{4}\delta^{ab}
\pi_{\mu\nu}\left(1+\delta^{ab}\frac{16}{\pi^{2}}\frac{g^{2}N}{p}\ln p+\text{finite}\right),
\end{equation}
which means that 
\begin{equation}
  \delta = \frac{8}{\pi^2} \frac{g^2 N}{p}.
\end{equation}

In principle, for the full calculation of the monopole non-Gaussianity, we would need to calculate the 3-point function 
at 2-loops.

Indeed, in the general formula for 3-point functions (\ref{3pointgeneral}) for 
$\Phi = A_\mu^a$  a bulk gauge field dual to a global symmetry current $\mathcal{O} = j^{a\mu}$ 
in the quantum field theory, we have
\bea
  &&  \langle A_{\mu}^a(p_{1})A_{\nu}^b(-p_{2})A_{\rho}^c(-p_{3})\rangle\cr
  &=& - \frac{1}{3} \frac{{\rm Re}
    \langle j^{d\lambda}(p_{1})j^{e\sigma}(-p_{2})j^{f\tau}(-p_{3})\rangle}{ {\rm Re}\langle j^{a\mu}(p_{1})j^{d\lambda}(-p_{1})\rangle 
    {\rm Re}\langle j^{b\nu}(p_{2})j^{e\sigma}(-p_{2})
    \rangle {\rm Re}\langle j^{c\rho}(p_{3})j^{f\tau}(-p_{3})\rangle}\;,\label{3pointAmu}
\eea
so we see that in order to get the full (anomalous) momentum dependence of the above 3-point function of $A_\mu^a$'s, we 
would need the full (anomalous) momentum dependence of the 3-point function of $j_\mu$'s, which is obtained first 
at 2-loops. 

As before, we must also do the Wick rotation $p\rightarrow \bar p=-ip$, $N^2\rightarrow -N^2$ before we apply to cosmology. 

However, as in \cite{Cravo:2023fqf}, we can at least make a shortcut, and observe that in a special momentum configuration we 
can write things completely in terms of 2-point functions. If we consider 
\be
p_1^2\ll p_2^2\simeq p_3^2\equiv p^2\;,\;\;\;
p^\mu\equiv p_2^\mu\simeq -p_3^\mu\;,\;\;\; p_1^\mu\simeq 0\;,\label{approx}
\ee
and with 
\be
t= 2\pi \frac{N^2}{4}p^{n_m-1}\;,\;\; n_m=1+2\delta,
\ee
we get
\bea
\langle j^a_\mu(p_1)j^b_\nu(-p_2)j^c_\rho(-p_3)\rangle&\simeq&
-\frac{1}{2}\epsilon^{ade}\langle j_\mu^d(p_2)j_\nu^b(-p_2)\rangle \frac{1}{\frac{N^2}{4} (d-1)p_3^{n_m-2}}
\frac{\d}{\d p_{3\sigma}}\langle j_\sigma^e (p_3)j_\rho^c(-p_3)\rangle \cr
&&+\frac{1}{2}\epsilon^{ade}\langle j_\mu^e (p_3)j_\rho^c (-p_3)\rangle\frac{1}{\frac{N^2}{4}
(d-1) p_2^{n_m-2}} \frac{\d}{\d p_{2\sigma}}
\langle j_\sigma^d(p_2)j_\nu^b(-p_2)\rangle.\cr&&\label{3pointspecial}
\eea

Substituting this 3-point function and the 2-point function in (\ref{3pointAmu}), and expressing things in terms of 
$p_1$ and $p$ in the approximation (\ref{approx}), we get
\bea
\langle A_{\mu}^a(p_{1})A_{\nu}^b(-p_{2})A_{\rho}^c(-p_{3})\rangle&\simeq& 
-\frac{\epsilon^{abc}}{6(N^2/4)^2}\frac{1}{{\rm Re}\;[p_1^{1+2\delta}]{\rm Re} \; [p^{1+2\delta}]}
\frac{\pi_{\lambda\sigma}(p)\frac{p_\tau}{p^2}
+\pi_{\lambda\tau}(p)\frac{p_\sigma}{p^2}}{\pi_{\mu\lambda}(p_1)\pi_{\nu \sigma}(p)\pi_{\rho\tau}(p)}\cr
&=&+\frac{\epsilon^{abc}}{6(\bar N^2/4)^2}\frac{1}{{\rm Im}\;[\bar p_1^{1+2\delta}]{\rm Im} \; [\bar p^{1+2\delta}]}
\frac{\pi_{\lambda\sigma}(\bar p)\frac{\bar p_\tau}{\bar p^2}
+\pi_{\lambda\tau}(\bar p)\frac{\bar p_\sigma}{\bar p^2}}{\pi_{\mu\lambda}(\bar p_1)\pi_{\nu \sigma}(\bar p)\pi_{\rho\tau}(\bar p)}.\cr
&&
\eea

This is the monopole non-Gaussianity that would be observed in the approximation (\ref{approx}) in the monopole distribution, 
in the case of the holographic cosmology.


\section{Conclusion}

In this note we clarified the calculation of $n$-point functions in holographic cosmology (of the McFadden-Skenderis type), 
by using the generalization of the Maldacena map from de Sitter inflation, based on the $\Psi_{\rm cosmo}[\phi]=Z_{\rm QFT}[\phi]$
equality and the resulting form (\ref{PsiPhi}) for the wavefunction of the Universe. After deriving general formulas, we have 
checked that indeed, the results for the 2-point functions of scalar and tensor fluctuations are obtained from 2-point 
functions of $T_{ij}$, as previously used in \cite{McFadden:2009fg,McFadden:2010na} based on direct computations in 
non-conformal holography. 

We then applied our results to the 2-point functions of gauge fields $A_i$, and showed how that is related to the resolution of the 
monopole problem, and calculated monopole non-Gaussianity, via the 3-point function of $A_i$. 

The methods used here are, however, general, so we can in principle use them to calculate any observable in holographic 
cosmology from $n$-point functions of operators in the 3 dimensional quantum field theory. So indeed, the Maldacena map can be 
used for a generic cosmology, mapped to a quantum field theory in one-dimension less, just like the equality of partition functions
($Z_{\rm CFT}=Z_{\rm gravity}$) in AdS/CFT was for generic gauge/gravity dualities, even non-conformal ones (and led 
to matching of correlators on both sides). 

\section*{Acknowledgements}

We would like to thank Kostas Skenderis for discussions.
The work of HN is supported in part by  CNPq grant 301491/2019-4 and FAPESP grant 2019/21281-4.
 HN would also like to thank the ICTP-SAIFR for their support through FAPESP grant 2021/14335-0.
The work of MC is supported by FAPESP grant 2022/02791-4.

\bibliography{wavefunction}
\bibliographystyle{utphys}

\newpage

\end{document}